\documentclass[aip,graphicx,twocolumn,letterpaper]{revtex4}
\usepackage{graphicx,amsmath,color}

 \newcommand{\m}[1]{\mathbf #1} 

\begin{document}

\title{Kinetic range spectral features of cross-helicity using MMS}

\author{Tulasi N. Parashar,  Alexandros Chasapis,  Riddhi Bandyopadhyay, 
Rohit Chhiber, W. H. Matthaeus, B. Maruca, M. A. Shay}
\affiliation{Bartol Research Institute, Department of Physics and Astronomy, University of Delaware, Newark, DE 19716, USA}
\author{J.~L. Burch}
\affiliation{Southwest Research Institute, San Antonio, TX, USA}

\author{T.~E. Moore}
\affiliation{NASA Goddard Space Flight Center, Greenbelt, MD, USA}

\author{C.~J. Pollock}
\affiliation{Denali Scientific, Fairbanks, Alaska, USA}

\author{B.~L. Giles}
\affiliation{NASA Goddard Space Flight Center, Greenbelt, MD, USA}

\author{D.~J. Gershman}
\affiliation{NASA Goddard Space Flight Center, Greenbelt, MD, USA}

\author{R.~B. Torbert}
\affiliation{University of New Hampshire, Durham, NH, USA}

\author{C.~T. Russell}
\affiliation{University of California, Los Angeles, CA, USA}

\author{R.~J. Strangeway}
\affiliation{University of California, Los Angeles, CA, USA}

\author {Vadim Roytershteyn}
\affiliation{Space Science Institute, Boulder, CO, USA}
\date{\today}

\begin{abstract}
We study spectral features of ion velocity and magnetic field correlations in 
the solar wind and in the magnetosheath using data from the 
Magnetospheric Multi-Scale (MMS) spacecraft. High resolution MMS
observations enable the study of transition of these correlations
between their magnetofluid character at larger scales into the 
sub-proton kinetic range, previously unstudied in spacecraft data. 
Cross-helicity, angular alignment and energy partitioning is 
examined over a suitable range of scales, employing measurements 
based on the Taylor frozen-in approximation as well as direct two-spacecraft correlation measurements. 
The results demonstrate signatures of alignment at large
scales. As kinetic scales are approached, the alignment between $\m{v}$
and $\m{b}$ is destroyed by demagnetization of protons.

\end{abstract}

\pacs{52.35.Hr,52.65.-y,96.50.Tf}

\maketitle 

{\bf \em Introduction}
Turbulence is a ubiquitous feature of astrophysical
plasma flows.  Interplanetary spacecraft observations 
have been used to study various aspects of plasma turbulence over the last
few decades including the systematic appearance of 
correlations of several types between fluctuations of the 
plasma 
velocity and the fluctuations of the magnetic field
(e.g. \cite{ColemanApJ68, BelcherDavis71, TuSSR95, BrunoLRSP13, ChenJPP16}
and many references therein). Such correlations are widely regarded as signatures 
of magnetohydrodynamic (MHD) fluctuations. Here we employ the unique observational capabilities 
of the Magnetospheric Multi-Scale (MMS) Mission
to examine 
these hitherto inaccessible
correlations at kinetic sub-proton scales. 

One of the features of MHD turbulence is the conservation of cross
helicity \cite{MatthaeusJGR82}. The normalized 
cross helicity is defined as 
\begin{equation}
\sigma_c = \langle |\delta z^+|^2-|\delta z^-|^2\rangle/\langle |\delta z^+|^2+|\delta z^-|^2 \rangle
\label{sceqn}
\end{equation}
where $\delta \m{z}^\pm = \delta \m{b}/\sqrt{\mu_0 m_i n_i} \pm \delta \m{v}_p$; 
$\delta \m{b}$ is the increment of
magnetic field fluctuation $\m{b}(\m{x})-\m{b}(\m{x}+\m{r})$, written 
in Alfv\'en speed units as suggested 
by this definition, where $\m{r}$ is a lag,  $n_i$ is the proton density, and %
$\m{v}_p$ is the proton fluid velocity fluctuation (mean removed), and the brackets indicate a suitable volume average.%
 $~\sigma_c$ is sensitive to 
both
the relative alignment of the velocity and magnetic
fluctuations, and their degree of 
energy equipartition. For an ``Alfv\'enic" state the magnitude of cross helicity
is very close to the value for Alfv\'en waves ($\sim 1$). A lower
value suggests a non-Alfv\'enic state. 
The importance of the ideal incompressible invariant
cross helicity $\langle |z^+|^2-|z^-|^2\rangle$
in selecting equilibria 
was noted early on by 
Chandrasekhar \cite{Chandrasekhar56}
and Woltjer \cite{Woltjer58a}.
Later, in the context of 
MHD turbulence theory  \cite{TingEA86,StriblingMatthaeus91}
it was noted that
relaxation of energy in both 2 and 3-dimensions
can lead to  
states that tend toward
{\it point-wise}
geometrical alignment, but not necessarily
equipartition, of $\bf v$ and $\bf b$. 
Simulations subsequently showed 
that this tendency occurs rapidly and locally 
\cite{MilanoEA01,MasonPRL06,MatthaeusPRL08,ServidioPP08}.
Subsequent research
exploited the tendency for 
alignment of $v$ \& $b$ fluctuations 
to estimate the degree of suppression of 
nonlinearity, thus influencing the 
slope of the turbulent spectrum 
\cite{BoldyrevPRL06,PodestaJGR09,PerezPRL09,PodestaApJ10-ch}. 

The situation becomes more complex
as proton kinetic scales
are approached,
and non-MHD effects are expected. 
As the
differential flows of protons and electrons become important,
the concept of helicity must be generalized
\cite{Turner86,ServidioPP08}. 
At sub-proton scales, one
could anticipate an electron MHD regime.  
However, in this paper we are interested in scales
approaching proton kinetic scales and hence limit ourselves to the
definition of $\sigma_c$ based on center of proton velocity.
The generalization of the 
cross helicity invariant \cite{ServidioPP08}
has not been discussed, as far as we know,
beyond a Hall MHD description. 
Even here, 
as kinetic scales are approached
the ``frozen-in'' approximation breaks down, 
small plasma
parcels are not tied to field lines and may not align with it.
Hence it is likely that flow-magnetic field alignment 
gets significantly modified.  However, how the alignment 
deviates from MHD behavior at 
kinetic scales has not been
studied, to the best of our knowledge.

The residual energy (difference of energy density 
in flow fluctuations and magnetic fluctuations)
is related in MHD to the alignment issue both kinematicaly and dynamically. 
The normalized residual energy, 
$\sigma_r = (\delta\m{v}^2 - \delta\m{b}^2) / (\delta\m{v}^2 +
\delta\m{b}^2)$
obeys an 
exact kinematic
relation involving the alignment cosine: 
$\cos( \theta ) \equiv 
(\delta \m{v} \cdot \delta \m{b}) /
\sqrt{|\delta \m{v}| |\delta \m{b}|} = \sigma_c/\sqrt{1-\sigma_r^2}$.
This relationship holds point-wise and for averages.
Even prior to invoking turbulence theory, one sees readily, for a pure non-dispersive 
Alfv\'en wave packet, that one necessarily has $\sigma_c = 1$, $\cos( \theta ) = \pm 1$, and $\sigma_r = 0$. 
Various turbulence theories pertain to the 
behavior of this angle, and to the residual energy,
in MHD and in the inertial range \cite{StriblingMatthaeus91,BoldyrevPRL06,MatthaeusPRL08}. To the best of our knowledge no theory exists that describes 
alignment in the kinetic range where the MHD approximation breaks down.

Here we use burst-mode data from the MMS spacecraft to provide a novel view of cross helicity, alignment and residual energy in the solar wind as well as the magnetosheath, 
exploiting MMS time-resolution and the multi-spacecraft
observations to probe these quantities in the kinetic range. 


{\bf \em Data:} 
We use data from the four spacecraft MMS mission, 
which emphasizes 
high cadence observations
of magnetic reconnection 
\cite{BurchScience16,BurchSSR16}. 
The magnetic field data are obtained from the Fluxgate Magnetometer
(FGM)  \cite{RussellSSR16},  providing
128 samples per second. 
The plasma moments used here 
are taken from the
Fast Plasma Instrument (FPI) \cite{PollockSSR16}, 
providing
proton moments at 15 samples per second, and electron moments at 30
samples per second. The four MMS spacecraft maneuver in a tetrahedral
formation, varying the inter-spacecraft separation over a wide range of
distances. The spacecraft separation during the intervals employed here 
lies 
below proton inertial length $d_i=c/\omega_{pi}$, where $c$ is the speed of light and $\omega_{pi}$ is the proton plasma frequency.
We study two intervals: i) A 40 minute long magnetosheath interval 
from 2017-12-26 starting at $06:12:43$ UTC
and ii) An hour long interval in the solar wind
from 2017-11-24 starting at $01:10:03$ UTC.

The solar wind data from FPI instrument was significantly polluted by
the spacecraft spin-tones and various measurement
 artifacts that are 
 more pronounced in the solar wind. The
 data is cleaned in two steps:
$\bullet$ The spin-tones and other spurious signals had distinctive
peaks in the spectrum. The peaks were identified by the Hampel algorithm
and their amplitudes were rescaled to the expected spectral amplitude
while keeping the phase information intact. $\bullet$ The resulting time
series was then low pass filtered at 1$Hz$ to 
avoid noise associated with the noise floor of the instrument.
For details of the cleaning algorithm, 
see \cite{BandyopadhyayApJ18}. 



\begin{figure}
\includegraphics[width=\columnwidth]{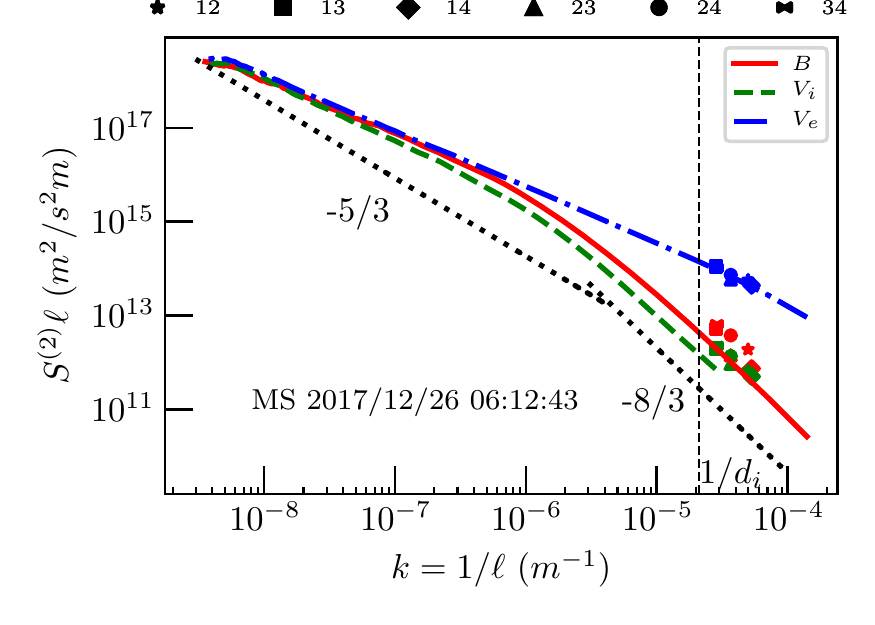}
\caption{Magnetosheath case. 
Equivalent spectra computed from structure functions show
the same behavior as the Fourier spectra. The spectra computed using
Taylor's hypothesis from single spacecraft measurements agree very well
with the multi-spacecraft estimates well into the kinetic range.}

\label{ms-bv-eqspec}
\end{figure}

{\bf \em Results: }
We begin by studying the spectral features of magnetic field and flows.
To leverage the advantages afforded by multi-spacecraft observations we
compute the equivalent spectra using the structure function technique as
described in \cite{ChasapisApJL17}. The second-order structure function
of a vector (e.g. magnetic field) is defined as

\begin{equation}
D_b^{(2)}(\m{r}) \equiv \left<|\m{b}(\m{x}+\m{r})-\m{b}(\m{x})|^2\right>
\end{equation}

With this definition of $D_b^{(2)}$, $S^{(2)}(\lambda) \equiv D^{(2)}(\lambda))\cdot\lambda$ behaves as
an ``equivalent spectrum'' as a function of an effective wavenumber
$k\equiv 1/\lambda$.

Structure functions from a single spacecraft use the Taylor hypothesis
to transform time lags into spatial lags, i.e., 
$\lambda = V \tau$ where $V$ is  the mean
flow speed and $\tau$ the time lag. 
Multi-spacecraft observations
enable the direct computation of 
structure function at a particular
physical lag $\lambda$ defined by the spacecraft separation, using time averaging to attain 
statistical significance. 
The four MMS spacecraft 
correspond to six different physical lags without resorting to Taylor's hypothesis. 
Given the small separation of the MMS spacecraft, the 
directly computed two point
structure functions can be compared
with the single spacecraft Taylor 
hypothesis results at scales well into the
kinetic range. 

\begin{figure}
\includegraphics[width=3.5in]{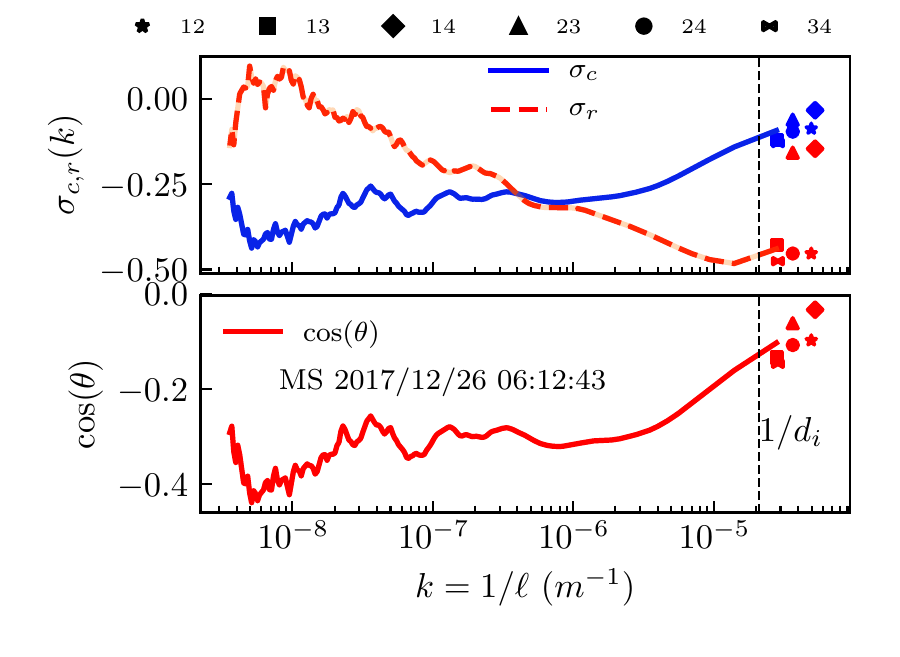}
\caption{Magnetosheath case:  
Equivalent spectrum of cross helicity. Solid line shows the equivalent spectrum from MMS 1, the
symbols show multi-spacecraft estimates. The legend at the top specifies symbols corresponding to 
the spacecraft pairs listed as numbers next to them.
At larger scales, the cross
helicity is $\sim -0.3$ and its magnitude decreases as approaching proton kinetic scales
and approaches zero deep in the kinetic range.}
\label{ms-sc-eqspec}
\end{figure}

Figure \ref{ms-bv-eqspec} shows the equivalent spectra for the magnetic
field, proton velocity, and electron velocity for the magnetosheath
interval of interest. 
As above, the magnetic field has been converted to Alfv\'enic units to make direct comparisons with  proton and electron velocities. In
the inertial range the magnetic field and the proton and electron
velocities have similar power.  However, they 
begin to depart from one
another at scales almost a decade 
larger than $d_i$.  In the kinetic range
the multi spacecraft observations match extremely well with the single
spacecraft observations computed using Taylor's hypothesis. This indicates
that Taylor's hypothesis is applicable even in the kinetic range at least
to scales slightly below the ion inertial length, consistent with earlier reports
\cite{ChasapisApJL17}. The distances between
spacecraft were spread over a 
moderate range (18.9-35.41 km) and the multi-spacecraft values follow the single spacecraft curve through this range.
The slight mismatch of single spacecraft and multi-spacecraft values
for protons is likely due to 
noise in proton measurements.

Having compared the single-spacecraft and multi-spacecraft observations
for the familiar spectra of the magnetic field and proton velocity, we
now proceed to compute the cross helicity spectrum using the structure
function technique described above. First the equivalent spectra of
$z^\pm$ are computed and Equation \ref{sceqn} is used to compute the
equivalent spectrum of cross helicity.

The top panel of figure \ref{ms-sc-eqspec} shows the equivalent spectrum for cross
helicity, computed from the Elsasser variables, for the magnetosheath interval. 
The figure also
shows the estimates of cross helicity at sub-proton lags computed
using multi-spacecraft lags. At larger scales the interval has a
cross helicity of -0.3 and it approaches 0 as we approach smaller
scales. Multi-spacecraft values not only match the single spacecraft
estimate, they continue the single spacecraft trend. The decrease in cross
helicity with decreasing lag indicates that the alignment between flow and
magnetic field decreases as kinetic scales are approached. The decrease
in cross helicity begins at about the scales where electron and proton spectra
depart from each other ($k\ell \sim 10^{-6}$), hence this decrease is
likely a direct result of break-down of the MHD approximation. The scales
between $k\ell \sim 10^{-6}$ and $k d_i =1$ are most likely described
by Hall-MHD physics, and the generalized helicity \cite{ServidioPP08}
would likely be better conserved at these scales. This however requires
a computation of vorticity using multi-spacecraft methods, and will be
considered in a future study.

The same panel shows the normalized residual energy spectrum for the magnetosheath case. $\sigma_r \sim 0$ at large scales, hinting at  
near-equipartition of energy between flow and magnetic energies. It decreases approaching smaller scales, indicating a loss in flow energy and dominance of magnetic energy at kinetic scales. Once again, a good agreement with single-spacecraft and multi-spacecraft values is observed. This hints at a difference between  magnetosheath and solar wind cases, as will be discussed below. 

The bottom panel of this figure shows the alignment angle as a function of scale. At large scales, it has a value $\sim -0.4$, indicating a degree of anti-alignment. Approaching kinetic scales, $\cos(\theta)$ approaches zero, indicating a lack of (anti-)alignment. This behavior is also 
seen for the solar wind intervals 
that we analyzed.

\begin{figure}
\includegraphics[width=3.5in]{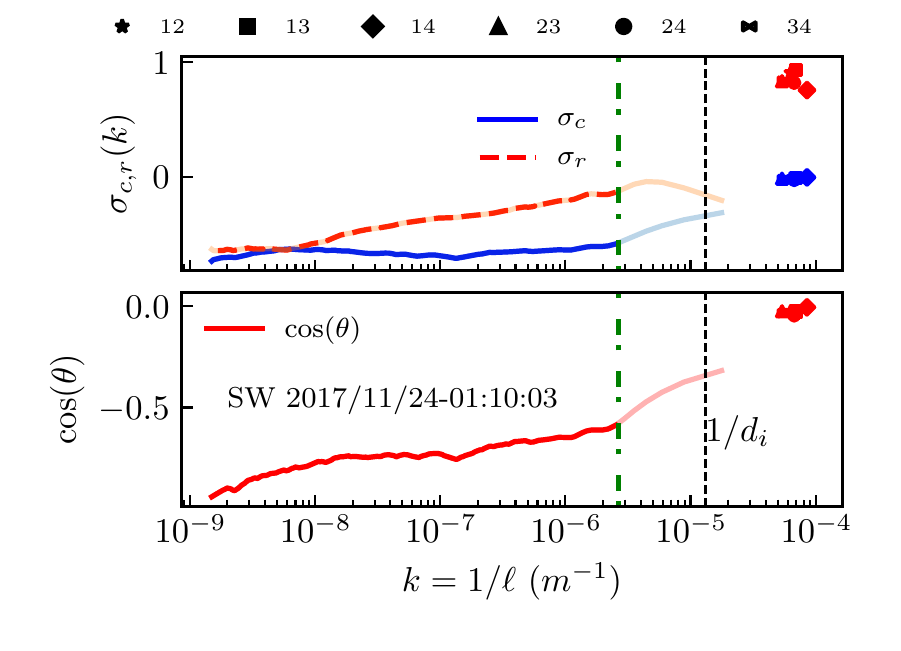}
\caption{Solar wind case: Equivalent spectrum of cross helicity. See text for details.}

\label{sw-sc-eqspec}
\end{figure}


We turn now to 
discussion of the spectral features of
the selected solar wind interval.
Spectra of magnetic field, proton flow and electron flow (not shown) 
display typical
features, e.g. power-law spectra, increasing kurtosis at smaller scales
etc.; see \citet{BandyopadhyayApJ18} 
for details including issues of cleaning FPI moments in the solar wind. 
Here
we directly move to the 
study of the equivalent spectrum of cross helicity
(shown in figure \ref{sw-sc-eqspec}). 
The vertical green dash-dotted line shows the
low-pass scale for the data cleaning 
procedure
\cite{BandyopadhyayApJ18}. The light blue cross helicity curve
extending beyond the low-pass cutoff is the ``Fourier interpolated'' data,
obtained at the original time cadence 
after imposing the low pass filter.
Although the results at these scales show a trend similar to
figure \ref{ms-sc-eqspec}, the results
in this range are of questionable 
validity.

\begin{figure}
\includegraphics[width=2.7in]{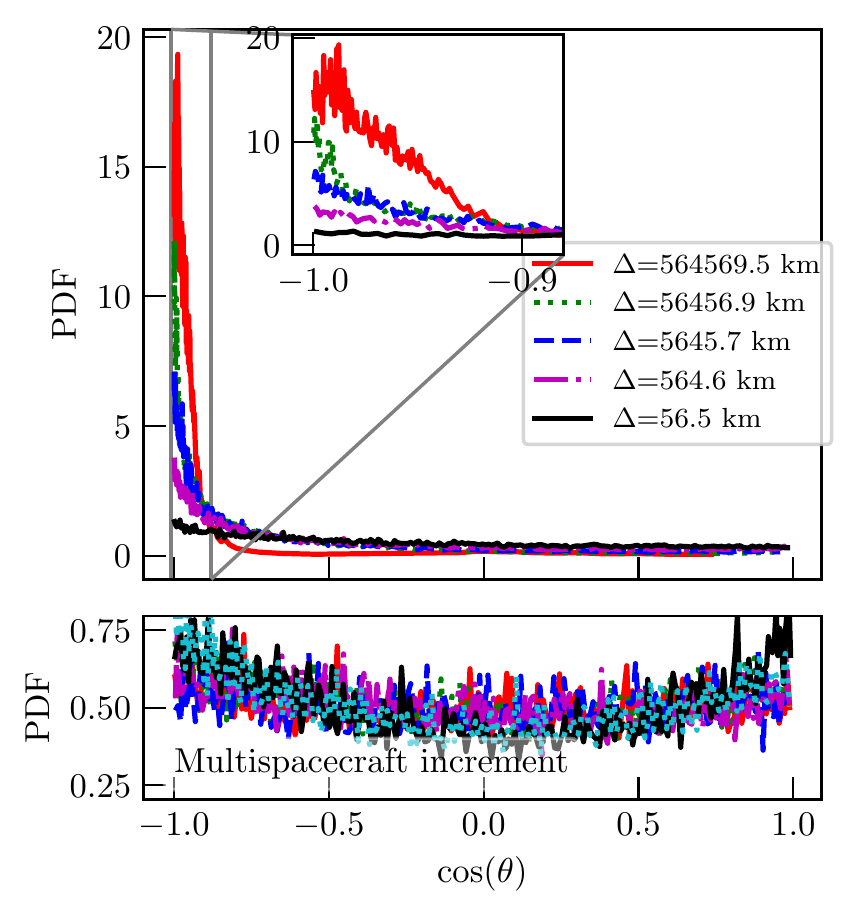}
\caption{Solar wind interval: 
PDFs of alignment angles for various increment lags (top panel) as well as multi-spacecraft lags (bottom panel). Different curves in the bottom panel correspond to different spacecraft pairs. The probability of high degree  of alignment decreases for smaller lags. 
See text for details. 
}
\label{sw-ctpdfs}
\end{figure}


The cross helicity in figure \ref{sw-sc-eqspec} remains close to -0.7 in
the inertial range. As kinetic scales are approached, the cross helicity
approaches zero, as was seen in the magnetosheath case. However, because
the data at these scales is interpolated, this conclusion is qualitative
at best. 
Nevertheless, this interpretation is 
further supported by multi-spacecraft 
measurements of $\sigma_c$, 
that are seen also to be close to zero. The overall qualitative picture is similar to what
was observed in the magnetosheath, supporting the idea that the trends
seen here are not an artifact of sampling, noise, or the data cleaning procedure.

The residual energy for the solar wind interval, 
also shown in Figure \ref{sw-sc-eqspec}, is magnetically dominated at large scales with a value $\sim -0.6$, 
and, moving towards proton scales, 
it approaches equipartition. 
It is interesting that
the multi-spacecraft estimates, deep 
in the sub-proton kinetic range, 
have values approaching $\sim 1.$,
indicating a preponderance of 
fluctuation flow energy compared
to magnetic energy. 
The alignment angle spectrum shown in bottom panel of
figure \ref{sw-sc-eqspec} also supports the ``isotropization'' conclusion. 
Evidently the multi-spacecraft data 
indicate that magnetic and velocity field have little or no preference to be aligned at the 10 km scale, deep
in the kinetic range. 

Alignment can be further studied by 
examination of the 
probability distribution
functions (PDFs) of alignment angles, shown in figure \ref{sw-ctpdfs}. 
The top
panel shows PDFs of $\cos( \theta )$
obtained from increments, from single spacecraft measurements and 
for lags ranging from 56 $km$ to 5.6 $\times$10$^6~km$, that is, from sub-proton scales to several correlation scales. 
The bottom panel
shows PDFs for $\delta\m{v},~\delta\m{b}$ alignment cosines computed from multi-spacecraft
lags, corresponding to kinetic scale measurements. The inset shows a magnification of
the initial part of the PDFs.  In this particular interval the alignment
PDFs peak at $\cos(\theta)=-1$, consistent with $\sigma_c=-0.7$ at large
scales. However, a counter-intuitive result is that decreasing the lag
decreases the alignment probability even in the inertial range. Once
kinetic scales are approached, the alignment is essentially absent,
consistent with demagnetization of the protons. Multi-spacecraft PDFs
show very slight departures from isotropy. However, the values are close
to 0.5 and the PDFs can be treated as almost isotropic.

\begin{figure}
\includegraphics[width=3.0in]{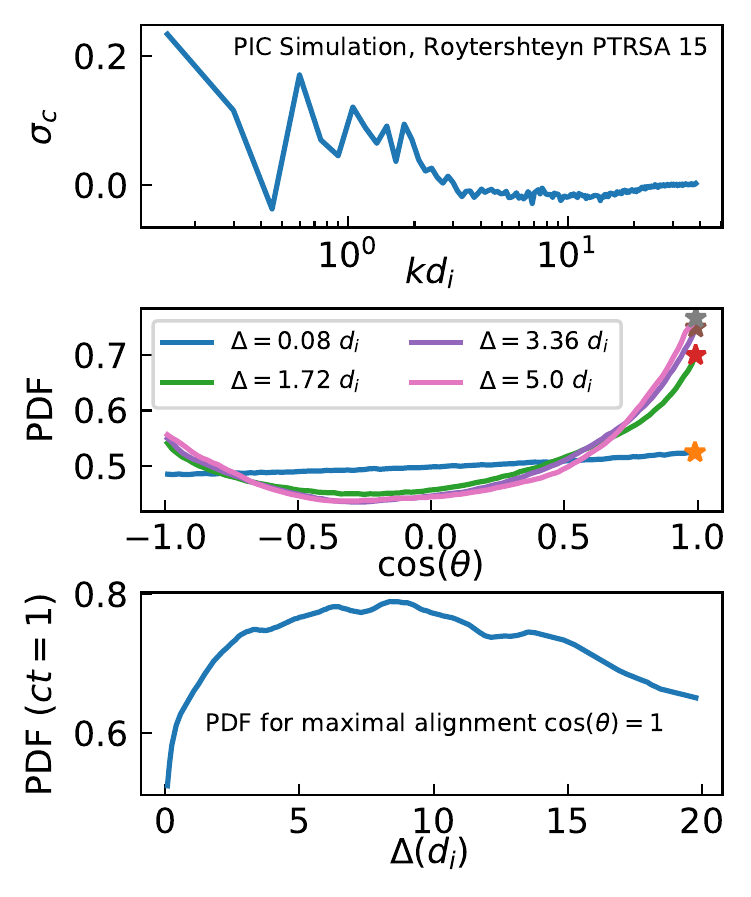}
\caption{Cross helicity spectrum, alignment PDFs for small increments, and probability of maximal alignment as a function of lag. See text for details.
}
\label{pic-results}
\end{figure}

{\bf \em Discussion: }
We have presented an analysis of 
proton velocity and magnetic field fluctuation correlations, alignment and partitioning of energy, studying the transition from MHD to kinetic scales.
Such studies are enabled in the MMS mission
by the unique combination of high time cadence and inter-spacecraft separation, 
in both magnetosheath and solar wind. 

Our main observational 
results, for the selected intervals,
are as follows:
$\bullet$
The normalized {\it cross helicity} $\sigma_c$
tends towards zero 
for decreasing scale approaching proton kinetic scales.
This has been anticipated in theory \cite{GrappinEA83}, but is not always manifest in observations \cite{TuEA90} nor
realized 
even in MHD for varying types of turbulence
(e.g., \cite{MattEA83}). 
Spacecraft observations of 
$\sigma_c$ in the kinetic range have not been previously reported. 
$\bullet$ The {\it residual energy}
$\sigma_r$ observed here is consistent with 
previous theoretical discussions (e.g., 
\cite{GrappinEA16} and Refs. within.) that argue for vanishing of $\sigma_r$ at high wavenumber in high Reynolds number MHD. 
Solar wind observations at MHD scales 
show a somewhat less clear tendency \cite{MarschTu90a}.
$\bullet$ The {\it alignment angle} 
also goes to zero towards kinetic scales
and into the kinetic range. This is a general phenomenon that we have observed in every interval that we analyzed, which is 
inconsistent with MHD simulation \cite{StriblingMatthaeus91} and MHD theory
\cite{BoldyrevPRL06}. Evidently it is a purely kinetic plasma physics phenomenon, 
deserving of further theoretical study.  



{\it Simulation.} 
To further support and elucidate 
these, we present an analysis 
from a fully kinetic 3D PIC simulation
\cite{RoytershteynPTRSA15}. The simulation was done on $2048^3$
grid points, with $L=42.~d_i$, $\beta_i=\beta_e=0.5$, with $\sim
2.6\times10^{12}$ particles, and an initial cross helicity of
$\sim 0.44$. The analysis is performed on a snapshot late in 
time-evolution of the simulation. For more details, 
refer to
\cite{RoytershteynPTRSA15}. Figure \ref{pic-results} shows the
corresponding simulation results for 
cross helicity, alignment, and residual energy $\sigma_c$, to be compared with the main observational
results above. 
The Figure shows
that $\sigma_c$ approaches zero in the kinetic range, consistent with the observations. 
The PDFs of $\cos(\theta)$ show decreasing 
probability of pure alignment approaching proton 
kinetic scales. 
In the bottom panel we also plot
the probability of maximal alignment 
(stars in middle panel) as
a function of lag. The probability of maximal alignment increases with
decreasing lag initially, 
consistent with MHD theories
\cite{BoldyrevPRL06, MasonPRL06, MatthaeusPRL08}. However, at around $10~
d_i$, the maximal probability begins 
to drop, indicating increasing frequency of  
non-aligned proton flow and magnetic field 
at smaller scales.

These results strengthen the idea that the alignment dynamics are much richer in the kinetic range than they are in inertial range dominated by MHD. The 
MMS instrument suite 
opens the doors
for these and other novel 
space plasma 
studies by providing high resolution plasma measurements in the kinetic range. Data sets with varied parameters and of longer durations are needed to start exploring some of these 
counter intuitive and ``non-MHD'' results. Generalizations of cross helicity, EMHD behavior deep in the kinetic range, as well as further 
implications of these concepts for kinetic range turbulence are topics 
for future study.

\acknowledgments
This study was supported in part by NASA by the MMS T\&M project under 
grant NNX14AC39G, by LWS program under 
grant NNX15AB88G, by 
the Heliophysics Guest Investigator program under 
grant NNX17AB79G, by NASA grant NNX17AI25G, and by NSF SHINE program under grant AGS-1460130 .

\end{document}